\documentclass[a4paper]{jpconf}
\usepackage{graphicx}

\usepackage{aas_macros}
\usepackage{amsmath}
\usepackage{amssymb}
\usepackage{hyperref}
\usepackage[sort&compress,numbers]{natbib}
\usepackage{subcaption}
\usepackage{xspace}

\usepackage{wrapfig}

\usepackage{color}
\begin{document}

\title{Modeling X-ray Bursting Neutron Star Atmospheres}

\author{L~Colleyn$^{1,2}$, Z~Medin$^{2}$, and A~C~Calder$^{1,3}$}

\address{$^1$ Department of Physics and Astronomy, 
Stony Brook University, Stony Brook, NY 11794-3800, USA}
\address{$^2$ Los Alamos National Laboratory, Los Alamos, NM 87545, USA}
\address{$^3$ Institute for Advanced Computational Science,
Stony Brook University, Stony Brook, NY 11794-5250, USA}
\ead{alan.calder@stonybrook.edu}

\begin{abstract}
We present a verification of a computational model, developed at the Los Alamos National Laboratory (LANL) for simulating radiation transfer in X-ray bursting neutron star atmospheres. We tested a baseline case and demonstrated strong agreement in the behavior of the outgoing spectrum's color-correction factor with earlier work and theoretical expectations.  By analyzing the relationship between the simulation time and outgoing flux, we also demonstrated how the model calculates through a sequence of time-independent atmospheric snapshots, each iteratively refined, and uses them to progressively converge toward the correct atmospheric state (as would be observed
during a burst). We examined the behavior of the outgoing flux across different optical depths and explored the physical explanations for deviations from a pure blackbody spectrum, attributed to frequency-dependent opacity sources. Additionally, we assessed the impact of Compton scattering, highlighting its role in redistributing photon energies. Finally, our analysis of spectra at various luminosities confirmed the expected trend of higher luminosities leading to a shift in peak emission and a larger overall radiation output. Ultimately, our findings verify the model's methods and motivate further exploration of different parameter spaces.
\end{abstract}

\section{Introduction}

%\subsection{Neutron Stars}

\hspace*{2em}Neutron stars, dense remnants of massive stars that have undergone a core collapse supernova, are among the most extreme stellar objects in the Universe. As endpoints of stellar evolution for stars with initial masses between $8–20 \text{ M}_\odot$, they are extraordinarily dense, with a radius around 10 km and masses typically between $1–2 \text{ M}_\odot$. These conditions make neutron stars unmatched astrophysical laboratories for studying dense matter physics and testing fundamental theories of nuclear matter.

Neutron stars form when a massive star exhausts its nuclear fuel. At the end of its life, a massive star has an inert,
degenerate iron core, and when the mass of this core reaches the maximum mass that can be supported by electron degeneracy, i.e.\ the Chandrasekhar limit, it collapses. The collapse halts when the density of the core approaches the nuclear saturation density ($n_s \approx 0.16 \; \: \text{nucleons/fm}^{3}$), and if conditions are right, this proto-neutron star will avoid collapsing to a black hole and a neutron star is born. 
A nice review of the formation, structure, internal composition and evolution of neutron stars may be found in \cite{Lattimer_2004}.

%\subsection{Constraining Neutron Star/Dense Matter Parameters}

A neutron star consists predominantly of neutrons, with smaller fractions of protons, electrons,
muons, and other matter. The star is supported primarily from the repulsive strong interaction, 
with neutron degeneracy pressure also contributing. Because the strong force is not well characterized, the cores of neutron stars may also harbor exotic states of matter, such as hyperons, kaons, pions, and deconfined quarks or ``strange quark matter" (SQM) \cite{Lattimer_2004}. Discerning the state of matter under these
extreme conditions is a challenge, as the material under these conditions is not experimentally
accessible.

Neutron star properties follow from the equation of state (EoS) for dense matter as it dictates the relationship between pressure and density, and the uncertainty in the strong force
leads to  epistemic uncertainty about the EoS at high densities. 
The EoS of dense matter establishes the mass-radius (M-R) relationship of neutron stars, with current models suggesting that maximum neutron star masses could reach up to about 2-2.5 $\text{ M}_\odot$, with radii around 10-12 km. 

Models of neutron
stars based on proposed EoSs can thus be confronted with observations for validation.
Not surprisingly, though, inferring masses and radii from observations can be challenging. 
Most neutron stars are observed as pulsars, allowing for mass determinations via measurement of general relativistic effects on Keplerian parameters. Radius measurements of neutron stars, however, show less precision -- the leading techniques being modeling of thermal emission from transiently quiescent neutron stars and isolated sources and neutron stars experiencing a thermonuclear runaway
of an accreted layer on the surface known as an X-Ray Burst (XRB). Information about the masses and radii of neutron stars can
also be obtained from accreting rapidly rotating pulsars. Pulsations from
accretion hot spots on the magnetic poles, offset from the rotation
axis, have profiles the shape of which are influenced by the
compactness of the neutron star. Analysis of observations of these pulse profiles thus yields constraints on masses and radii \cite{Ozeletal2016,Salmietal2018,Bogdanovetal2019}. Also, although beyond the scope of this work, we note that information
about neutron star masses and radii can be inferred from the observation of gravitational radiation from neutron star mergers. See \cite{FieldsRadice2025} and references therein for more information. 

Generally, X-ray bursts (XRBs), transient events observed from neutron stars in Low-Mass X-ray Binary (LMXB) systems, offer a unique window into their extreme environments. First observed in 1975 by NASA's SAS-3 X-ray satellite \cite{Mayer1975}, XRBs originate when a neutron star accretes typically hydrogen- and helium-rich material from a low-mass companion. The accreted material forms a layer on the neutron star's surface, where intense gravitational compression raises the temperature and density. Thermonuclear fusion ignites in this layer, leading to a thermonuclear runaway that releases energy as a burst of X-rays. This energy, initially generated deep within the envelope, interacts with the photosphere, producing a spectrum well-approximated by a modified black-body distribution. This burst spectrum is ``modified" by absorption and scattering processes in the atmosphere, shifting the observed spectrum to a ``color temperature" different from the effective temperature of the photosphere. This logic applies  
to one special case of XRBs as well, Photospheric Radius Expansion (PRE) bursts \cite{Steiner_2013}.

Spectral modeling of emission from neutron star atmospheres can yield mass, radius, effective temperature, and surface gravity constraints. In PRE bursts, a thermonuclear explosion on the surface generates intense luminosity that reaches the Eddington limit, causing the outer layers of the star’s atmosphere to expand. As the burst subsides, the atmosphere contracts back to the neutron star’s surface, reaching the ``touchdown" point, where the photosphere settles at the star’s surface. By analyzing the spectral evolution throughout this process, particularly at the peak flux and touchdown point, we can infer neutron star masses and radii, provided we have an independent measurement of the source distance or the outgoing flux at the burst peak \cite{inproceedings}. 

Accordingly, XRB simulation models are crucial for bridging the gap between observations and theoretical models. Simulating XRBs requires resolving complex nuclear reactions and radiative processes on timescales less than a nanosecond, making full-scale modeling computationally challenging. Some current efforts rely on one-dimensional simulations with numerous other approximations to reduce complexity, but can account for the atmosphere's composition, surface gravity, and radiative transfer processes. 

We present results from such an XRB modeling tool, named ``ZCODE."  ZCODE was developed at LANL and is first described in \cite{Medin_2016}. This model aims to address the uncertainties inherent in XRB simulations by incorporating detailed physics, including atmospheric composition, surface gravity, and radiative transfer processes. Our main objective in this work is first to verify the consistency of the results from simulations using this model with those of similar models. By doing so, we contribute to refining the theoretical framework needed to interpret observational data and deepen our understanding of neutron star properties.

\section{Equations}

\hspace*{2em}The state of the neutron star atmosphere is modeled through three main equations: The radiation transfer equation, which describes the movement of the radiation field from the interior to the surface as it exchanges energy with the atmosphere's material, the material energy equation, which describes radiation-material energy exchange from the material's side, and the hydrostatic balance equation, which describes the density distribution of the atmosphere in equilibrium when there is no bulk radial motion in the material due to the balance between the gravitational and pressure forces. A complete description of the equations used in our model is found in \cite{Medin_2016}; here we outline the basic structure.

%\subsection{Radiation Transfer Equation}

Ignoring general relativistic effects and bulk motion of the photosphere (but see \cite{Medin_2016}), the equation of radiation transfer is given by:
\begin{equation}
    \begin{aligned}
         \frac{1}{c}\frac{\partial I_\nu(\hat{\Omega})}{\partial t} + \hat{\Omega} \cdot \nabla I_\nu(\hat{\Omega}) = - \rho \kappa^{\rm tot}_\nu(\hat{\Omega}) I_\nu(\hat{\Omega}) + j^{\rm tot}_\nu(\hat{\Omega}).
    \end{aligned}
    \label{EqRadTrans}
\end{equation}
Here, $c$ is the speed of light, $\nabla$ is the spatial gradient, $\rho$ is the material density, and $I_\nu(\hat{\Omega})$ is the specific intensity at photon frequency $\nu$ in propagation direction $\hat{\Omega}$. Additionally, $\kappa^{\rm tot}_\nu(\hat{\Omega}) = \kappa_\nu(\hat{\Omega}) + \kappa^{\rm sc}_\nu(\hat{\Omega})$ is the total opacity with $\kappa_\nu$ the absorption opacity and $\kappa_\nu^{\rm sc}$ the scattering opacity; and $j^{\rm tot}_\nu(\hat{\Omega}) = j_\nu +j^{\rm sc}_\nu(\hat{\Omega})$ is the total emissivity with $j_\nu$ the isotropic emission coefficient and $j_\nu^{\rm sc}$ the scattering coefficient. We also note that the intensity and the various coefficients depend on time ($t$) and position ($r$), although not stated explicitly.

Local thermodynamic equilibrium is assumed everywhere in the atmosphere, such that the absorption opacity and emission coefficient are related via Kirchoff's law of thermal radiation. 
Complete details of the opacities and the calculation thereof
may be found in \cite{Medin_2016}.
We note that the publicly accessible TOPS Opacities webpage \cite{TOPS_Opacities}, powered by the LANL OPLIB database, is used to obtain the absorption opacity as a function of $\rho$ and material temperature $T$.

%\subsection{Material Energy Equation}

Assuming infinite ion-electron coupling (ions and electrons are always in thermal equilibrium) and ignoring heat conduction, the material energy equation is given by:
\begin{equation}
    \begin{aligned}
         \rho c_V \frac{\partial T}{\partial t} = \int_0^\infty d\nu \int_{4\pi}d\Omega [\rho \kappa_\nu^{\rm tot}(\hat{\Omega})I_\nu(\hat{\Omega}) - j_\nu^{\rm tot}(\hat{\Omega})].
    \end{aligned}
    \label{EqMatE}
\end{equation}
Here, the total (ion + electron) specific heat $c_V$ is calculated assuming the material is an ideal gas.

%\subsection{Hydrostatic Balance Equation}

Finally, the hydrostatic balance equation is given by:
\begin{equation}
    \begin{aligned}
         \frac{\partial P_{\rm gas}}{\partial r} = - \frac{1}{c} \int_0^{\infty} d\nu \int_{4\pi} d\Omega \mu[-\rho \kappa_\nu^{\rm tot}(\hat{\Omega})I_\nu(\hat{\Omega}) + j_\nu^{\rm tot}(\hat{\Omega})] - \rho g.
    \end{aligned}
    \label{EqHydro}
\end{equation}
Here, $g = (1+z)GM/R^2$ is the surface gravitational acceleration, with $G$ the gravitational constant, $M$ the neutron star mass, $R$ the neutron star radius, and $z$ the gravitational redshift at the surface of the star (but see \cite{Medin_2016}, where $g$ and $z$ are allowed to change with photosphere height). As with the specific heat above, the gas pressure $P_{\rm gas}$ is calculated assuming an ideal gas EoS.

\section{Simulation Methods}

%\subsection{General Model and Initial Conditions}

\hspace*{2em} In equilibrium, our model (detailed in \cite{Medin_2016}) is fully determined by the chemical composition of the atmosphere $\{X\}$ (assumed to be uniform and constant in time for simplicity), $g$ and $R$ (which determine $M$), and the surface luminosity $L$. However, since we cannot directly set the luminosity of our model, we use the temperature at the base of the atmosphere $T_{\rm base}$ as a proxy. To obtain an equilibrium atmosphere and the equilibrium outgoing radiation spectrum, we fix $\{X\}$, $g$ and $R$ and adjust $T_{\rm base}$ in an iterative fashion: we make an initial guess for $T_{\rm base}$ and run our model to determine the resulting atmosphere state and the surface luminosity; and then gradually adjust $T_{\rm base}$ while calculating the new state, until the desired value of $L$ is reached. Below we outline our model; a full description of the methodology used and a detailed explanation for the assumptions made can be found in \cite{Medin_2016}.

The atmosphere model consists of 100 discrete cells arranged in a one-dimensional grid; each cell represents a spherical shell of the atmosphere and the shells are equally
%\begin{figure*}[h]
%\includegraphics[width=20pc]{diagram.png}
%\begin{minipage}[b]{17pc}\caption{\label{Diagram} 
% Simplified schematic depicting the atmosphere's cell structure as spherical shells of the atmosphere, with material quantities defined within each discrete cell and at the base.
%    \label{Fig1}
%}
% \end{minipage}
%\end{figure*}
\begin{wrapfigure}{r}{7.0cm}
\centering
\includegraphics[width=6cm]{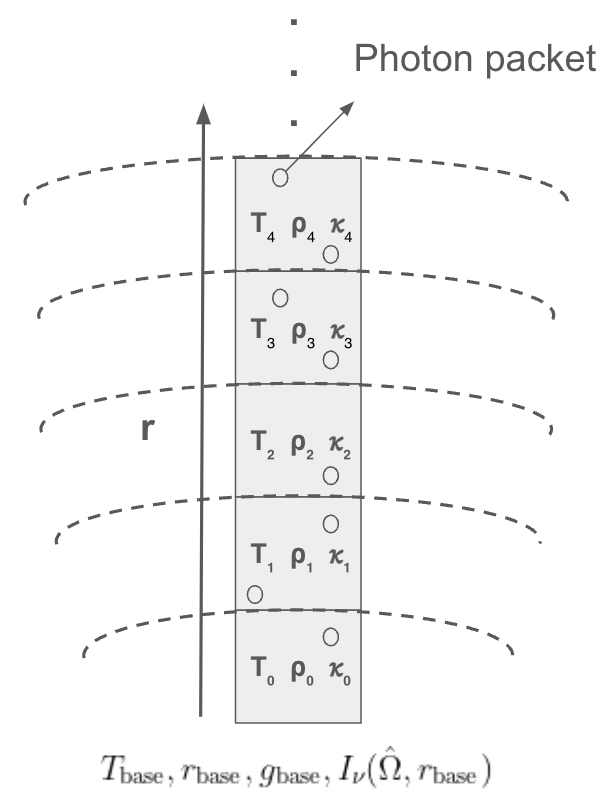}
%\begin{minipage}[b]{17pc}
\caption{\label{Diagram} 
 Schematic depicting the atmosphere's spherical cell structure with material quantities defined within each discrete cell and at the base. Note that in the terminology of this section, $r_{\rm base} = R$, and $g_{\rm base} = g$ is assumed constant.
    \label{Fig1}
}
% \end{minipage}
\end{wrapfigure}
spaced in radius. Density, temperature, and radiation intensity values are defined at the centers each cell. This arrangement is shown in Fig.\ \ref{Diagram}.
The atmosphere extends to densities of a few $\rm{g}/\rm{cm}^3$. We assume that the radiation at the base of the atmosphere is in thermal equilibrium with the material, such that the intensity there is described by the Planck function:
\begin{equation}
    \begin{aligned}
       I_{\nu,{\rm base}}(\hat{\Omega}) = B_\nu(T_{\rm base}).
    \end{aligned}
    \label{EqIbase}
\end{equation}
We utilize opacity values covering a broad range of photon energies, spanning 300 logarithmically spaced frequency groups between $\nu = 10~\rm{eV}$ and $\nu = 1~\rm{MeV}$.

After picking the desired values for $\{X\}$, $g$ and $R$, and the target value for $L$, the next step is to establish an initial guess for $T_{\rm base}$ and for $\rho$, $T$, and $I_\nu(\hat{\Omega})$ everywhere in the atmosphere. In principle, the model can converge to an equilibrium solution regardless of the initial estimates of density, temperature, and radiation intensity. However, choosing initial conditions close to the final equilibrium state is advantageous, as it minimizes numerical instabilities and reduces the number of iterations required for convergence. Accordingly,
we implement a model that balances ease of implementation and fast initialization with accuracy, using several simplifying assumptions. First, using the opacity profile, we guess the initial temperature profile of the atmosphere based on the ``Eddington + grey atmosphere" approximation, which relates optical depth ($\tau$) and temperature via:
\begin{equation}
    \begin{aligned}
      [T(\tau)]^4 = T^4_{\rm eff} \frac{3}{4}\left(\tau + \frac{2}{3}\right),
    \end{aligned}
    \label{EqTprof}
\end{equation}
where the effective temperature at the surface $T_{\rm eff} = [L/(4\pi R^2 \sigma_{\rm SB})]^{1/4}$ is given by the Stefan-Boltzmann law.
Second, the initial guess for the mass density profile is directly determined by the temperature profile, such that hydrostatic balance is satisfied in each cell. Third, the density profile is used to calculate the optical depth profile, and then the radiation field in each cell is given by a Planck function at the temperature calculated from Eq.\ (\ref{EqTprof}).

%\subsection{Monte Carlo Method}

Our main equations of radiation transfer and material energy are solved using an Implicit Monte Carlo method. In this method, the material properties are determined only at cell centers and updated only at the end of specified time intervals, called cycles (and the cycle number is the number of such intervals that have elapsed since the beginning of the simulation).
On the other hand, the radiation field evolves continuously in time and space, through the propagation of many discrete, stochastic particles. These Monte Carlo particles can be created at, move to, and be destroyed at any location in the grid, at any time, based on physics-weighted random chance. Since the material properties are only updated at cell centers and ends of cycles, the depositing and removal of material energy from a cell must be handled in a semi-implicit manner to avoid numerical issues like negative cell energies (see \cite{Medin_2016} and references therein).

Ideally, a Monte Carlo particle can be created at any time and location in the simulation. However, this is not always possible due to computational considerations: if the number of existing particles is too large, a new one might not be created. To account for this, each Monte Carlo particle carries a weight. During times when the number of existing particles is high, few new particles will be created, but they will have large weights because they represent many photons. Particle creation is chosen in a weighted-random manner, with high-density and high-temperature regions more likely to emit high-weight particles; in addition, particles are more likely to be emitted with frequencies near the peak of the blackbody corresponding to the local temperature. Particles are also created at the inner boundary of the simulation, representing the escape of photons from the hot layers below the atmosphere.

After creation, each particle will travel in a straight line until it undergoes an event, which could be absorption, scattering, crossing a cell boundary, or reaching the end of a cycle. The code calculates a distance associated with each possible event, determining which will occur first. Particles that exit through the inner boundary are destroyed, while those leaving through the outer boundary have their energy weight recorded before being destroyed. At the end of each cycle, the energy in each cell is updated based on the combined energy weights of all particles created in that cell (emission), destroyed in the cell (absorption), and modified (up- or down-scattering).

This process of transporting photons and coupling them to the material continues until steady-state and radiative equilibrium is reached. The criterion for reaching the steady-state is that $L$ changes by less than one part in $10^6$ per cycle, at which point the atmosphere is very nearly in radiative equilibrium.
The atmosphere is adjusted to maintain hydrostatic balance every several cycles by modifying the mass density in each atmospheric cell to satisfy the equation of hydrostatic balance, while keeping the density at the base of the atmosphere fixed.

At the end the procedure, the simulation has reached a steady-state; however it has not necessarily reached the correct steady-state. In each of the atmosphere simulations described in our results (next section), we require a specific value for the surface luminosity. Due to the approximate nature of our initial guess for the temperature profile (Eq.\ \ref{EqTprof}), we do not simply choose $T_{\rm_{base}}$ to give us that $L$ value. Instead, we must iterate through the steps described in this section, gradually adjusting the base temperature until the target $L$ value is reached.

\section{Results}

\hspace*{2em}All displayed results are from atmospheres composed purely of hydrogen, with $g = 10^{14} \: \rm{cm} \; \rm{s^{-2}}$, and $R = 11.5 \: \rm{km}$. Our primary goal is to verify the model's methodology, ensuring it performs correctly for this baseline case, providing a foundation for exploring new setups (different atmospheric compositions, surface gravities, etc.) in future work. With that in mind, in this section we present results where we instead varied the minimum optical depth in the simulation ($\tau_{\rm min}$), the maximum number of Monte Carlo particles in the simulation ($N_{\rm tot}$), or the luminosity ratio ($l$), the last defined as:
\begin{equation}
    \begin{aligned}
       l = \frac{L}{L_{\rm T}},
    \end{aligned}
    \label{EqLratio}
\end{equation}
with $L_{\rm T}$ the ``Thomson" Eddington luminosity $L_{\rm T} = (1+z)4 \pi c G M/\kappa_{\rm T}$ and $\kappa_{\rm T}$ the Thomson scattering opacity. To test the quality of our model, we study the convergence of our results with varying $\tau_{\rm min}$ and $N_{\rm tot}$, analyze the effect of $l$ on the spectrum, and compare our results with those of other works (\cite{Suleimanov_2012}).

\subsection{Total Outgoing Flux}

\begin{figure}[htpbp]
    \centering
    \begin{subfigure}{0.49\textwidth}
        \centering
        \includegraphics[width=\linewidth]{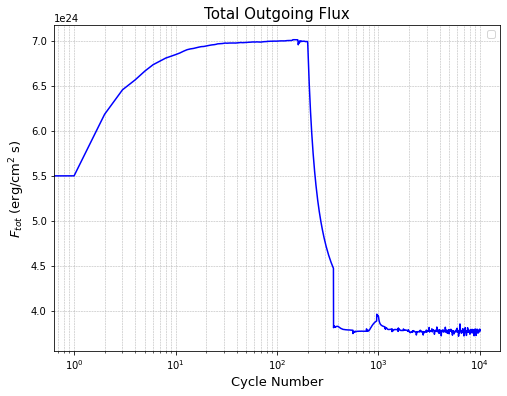}
        \caption{}
        \label{fig:fig2a}
    \end{subfigure}
    \hfill
    \begin{subfigure}{0.49\textwidth}
        \centering
        \includegraphics[width=\linewidth]{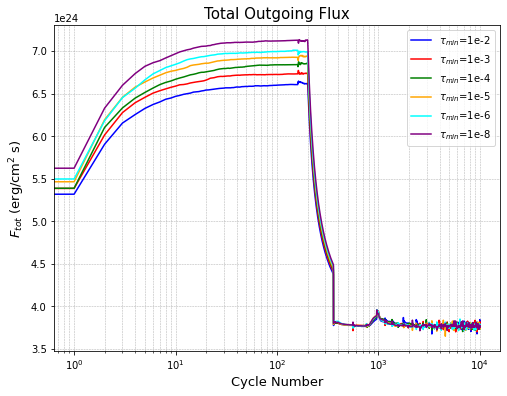}
        \caption{}
        \label{fig:fig2b}
    \end{subfigure}

    \caption{ Total outgoing flux vs.\ cycle number at $l = 0.5$
    with the fiducial value of $\tau_{\rm{min}}$ (left
    panel) and total outgoing flux vs.\ cycle number at $l = 0.5$ with varying values of $\tau_{\rm{min}}$.}
    \label{Fig2}
\end{figure}

\begin{figure}[htpbp]
    \centering
    \begin{subfigure}{0.49\textwidth}
        \centering
        \includegraphics[width=\linewidth]{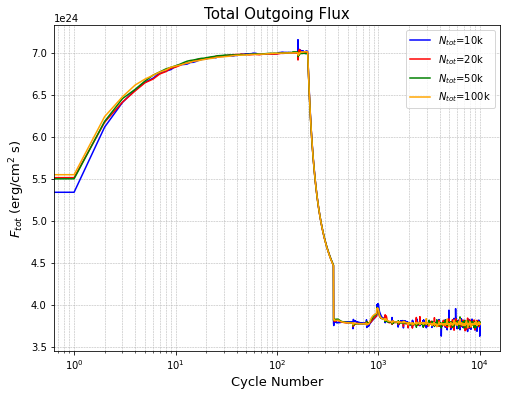}
        \caption{}
        \label{fig:fig2aa}
    \end{subfigure}
    \hfill
    \begin{subfigure}{0.49\textwidth}
        \centering
        \includegraphics[width=\linewidth]{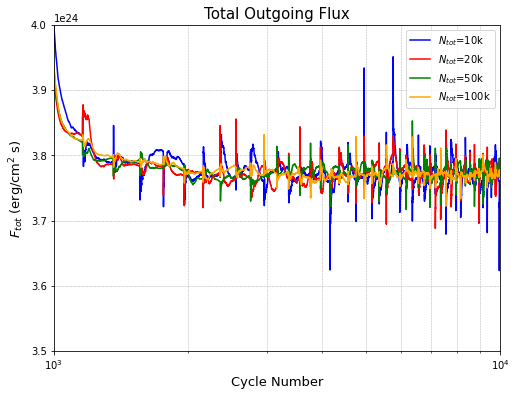}
        \caption{}
        \label{fig:fig2ab}
    \end{subfigure}

    \caption{Total outgoing flux vs.\ cycle number at $l = 0.5$ showing
    simulations with different numbers of photon packets. The panel on the right
    is detailed view of the results at large cycle numbers.}
    \label{Fig2a}
\end{figure}

\hspace*{2em}We analyze the evolution of the total outgoing flux $F_{\rm tot} = L/(4\pi R^2)$ as a function of the cycle number, as shown in Figures \ref{Fig2} and \ref{Fig2a}. In the simulations shown in Figures \ref{Fig2} and \ref{Fig2a}, the density profile is adjusted every 5 cycles and the base temperature every 200 cycles. Initially, the flux increases steadily, gradually rising toward a peak that represents the maximum radiation output of the system. This phase is followed by a sharp and sudden drop in flux at 200 cycles, which corresponds to the first and largest code adjustment of the atmosphere's temperature at the base. After this abrupt decline, the flux stabilizes as the system moves toward its equilibrium state; the flux continues to exhibit small fluctuations due to the hydrostatic re-balancing of the density every several cycles.
%The logarithmic scale on the x-axis of  Figures \ref{Fig2} and \ref{Fig2a} shows the rapid evolution: a rapid increase in flux for about 300 cycles, then a sudden drop in flux for about 200 cycles, followed by the longer-term stabilization phase of  roughly 9000 cycles. 
%\begin{figure}[htbp]
%    \centering
%    \includegraphics[width=\linewidth]{Ftot.png}
%    \caption{Total outgoing flux vs. Cycle number at $l_{\rm{proj}} = 0.5$.}
%    \label{Fig1}
%\end{figure}

Fig.\ \ref{Fig2} shows the total outgoing flux as a function of cycle number for $\tau_{\rm min} = 10^{-6}$ (the default value; left panel), and for a range of values (right panel). As can be seen from this figure, the value of $\tau_{\rm min}$ has a strong effect on the shape of the flux profile at early times, due to its influence on our initial guess for the atmosphere conditions; but it has almost no effect on the long-term evolution of the flux. In addition, once convergence is reached the peak of the outgoing radiation spectrum (i.e., the flux as a function of frequency) changes by less than 1\% across $\tau_{\rm min}$ values (trend not shown). These two results show that our model is robust to varying $\tau_{\rm min}$ -- effectively, after the density and temperature adjustment steps described above the atmosphere is completely reconfigured and loses all memory of its initial conditions.

Fig.\ \ref{Fig2a} shows the outgoing flux as a function of cycle number for different values of $N_{\rm tot}$. The left panel show the overall flux with cycle number and the right shows a detailed view of the converged region of cycle numbers. The
flux adjustments throughout the simulation cycles are smoother as $N_{\rm tot}$ goes up, particularly in the last half of the simulation as seen in the detailed view. While the computational time per cycle, per MPI
process, is about a factor of 10 larger for simulations with $N_{\rm tot} = 1e5$ than for simulations with $N_{\rm tot} = 1e4$, this study suggests that the increased number of particles is worth the added computational cost, due to the greatly improved convergence to the desired luminosity value.

%\begin{figure}[htpbp]
%    \centering
%    \includegraphics[width=\linewidth]{Ftot_Taulo.png}
%    \caption{Total outgoing flux vs. Cycle number at $l_{\rm{proj}} = 0.5$ with varying values of $\tau_{\rm{min}}$.}
%    \label{Fig2}
%\end{figure}

\subsection{Outgoing Spectrum}

\begin{figure}[htpbp]
    \centering
    \begin{subfigure}{0.49\textwidth}
        \centering
        \includegraphics[width=\linewidth]{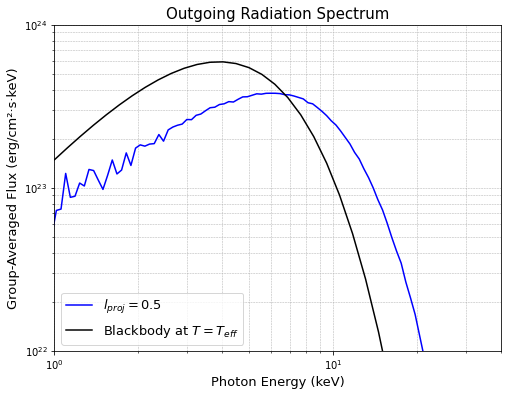}
        \caption{}
        \label{fig:Fig45combineda}
    \end{subfigure}
    \hfill
    \begin{subfigure}{0.49\textwidth}
        \centering
        \includegraphics[width=\linewidth]{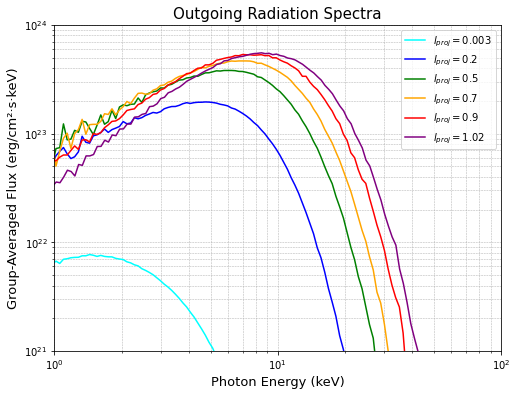}
        \caption{}
        \label{fig:Fig45compbinedb}
    \end{subfigure}

    \caption{Outgoing radiation spectrum at $l = 0.5$ (blue curve) with the blackbody approximation at the atmosphere effective temperature (black curve) (left panel).
    Outgoing radiation spectra at different luminosities, ranging from $l = 0.003$ to $l = 1.02$ (right panel). Note that in the panel labels the luminosity ratio is referred to as $l_{\rm proj}$; the added subscript there refers to the luminosity ``projected" on to the base of the atmosphere, as described in \cite{Medin_2016}.
    }
    \label{Fig45combined}
    
\end{figure}

\hspace*{2em} The equilibrium outgoing radiation spectrum, as shown in 
the left panel of Fig.\ \ref{Fig45combined}, deviates from a perfect blackbody distribution and is well-approximated by a ``diluted” blackbody distribution \cite{Medin_2016}. The spectrum exhibits both a noticeable blueshift and broadening, with a flatter distribution across a wider range of photon energies, compared to the blackbody at the effective atmospheric temperature $T_{\rm{eff}}$. This occurs because the photon energies are redistributed, with an excess of high-frequency photons and a suppression at lower frequencies relative to the standard Planckian distribution. This ``hardening" of the spectrum is expected from electron scattering in the atmosphere, and these results quantify the change in the emergent spectrum from our initial guess of a purely thermal equilibrium distribution.

%\begin{figure}[htpbp]
%    \centering
%    \includegraphics[width=\linewidth]{output.png}
%    \caption{Outgoing radiation spectrum for at $l_{\rm{proj}} = 0.5$ (blue line). The Blackbody approximation at the atmosphere effective temperature is shown as a black line.}
%    \label{Fig3}
%\end{figure}

Comparing spectra at varying luminosity values, Fig. \ref{Fig45combined} 
(right panel) shows the outgoing radiation spectra as flux versus photon frequency for different luminosities. As the luminosity increases, the spectrum shifts toward higher photon frequencies, and the overall flux rises; in addition, the spectrum narrows with increasing luminosity, leading to a more blackbody-like shape.
Note that the spectra in Fig.\ \ref{Fig45combined} are noisy at low photon energies. This is because there are fewer Monte Carlo particles contributing to the outgoing spectrum at these frequencies, leading to greater statistical fluctuation in flux. Increasing $N_{\rm tot}$ helps smooth the curves in this region, but does not affect the overall spectral shape or the dynamics of the simulation (as the number of actual photons in this region is very small).

%\begin{figure}[htpbp]
%    \centering
%    \includegraphics[width=\linewidth]{spectra.png}
%    \caption{Outgoing radiation spectra at different luminosities, ranging from $l_{\rm{proj}} = 0.003$ to $l_{\rm{proj}} = 1.02$. }
%    \label{Fig4}
%\end{figure}

%\begin{table}[htbp]
%\centering
%\begin{tabular}{|c|c|}
%\hline
%\textbf{$l_{\text{proj}}$ [L$_{\text{proj}}$/L$_{\text{T}}$]} & \textbf{$F_{\text{tot}}[\text{erg}/\text{cm}^2$ $\cdot$ \text{s}]}\\
%\hline
%0.003 & 2.222e+22 \\
%0.2   & 1.479e+24 \\
%0.5   & 3.713e+24 \\
%0.7   & 5.160e+24 \\
%0.9   & 6.657e+24 \\
%1.02  & 7.570e+24 \\
%\hline
%\end{tabular}
%\caption{Total radiation output ($F_{\text{tot}}$) for different values of $l_{\text{proj}}$.}
%\label{tab:flux}
%\end{table}
%
%We also tested the simulation under two different scenarios: one considering only Thomson scattering, and another incorporating Compton scattering alongside Thomson scattering. Fig.\ \ref{Fig5} compares the outgoing spectra for these cases. We observe that including Compton scattering results in a higher peak flux, while the case without Compton scattering exhibits a broader frequency range. Additionally, the spectrum in the full Compton scattering case again appears more jagged at lower photon energies.
%\begin{figure}[htpbp]
%    \centering
%    \includegraphics[width=\linewidth]{Scttrng_spectra.png}
%    \caption{Outgoing radiation spectrum at $l_{\rm{proj}} = 0.5$ considering both full scattering treatment with Compton scattering (blue line), and without Compton scattering (red line).}
%    \label{Fig5}
%\end{figure}

\subsection{Color-correction Factors}

\hspace*{2em}An essential aspect of modeling X-ray bursts is characterizing the outgoing radiation spectrum, allowing for reliable comparison with observed spectra. The color temperature $T_c$, the temperature of the blackbody curve which best matches the observed spectrum, is found by fitting the outgoing spectrum to a Planck function \cite{Medin_2016}.
We write this as $F_\nu \approx w B_\nu(T_c)$ with a perfect fit corresponding to $w = f_c^{-4}$, where $f_c = T_c/T_{\rm {eff}}$ is the color correction factor (or spectral hardness factor). 

\begin{figure}[htpbp]
    \centering
    \includegraphics[width=0.9\linewidth]{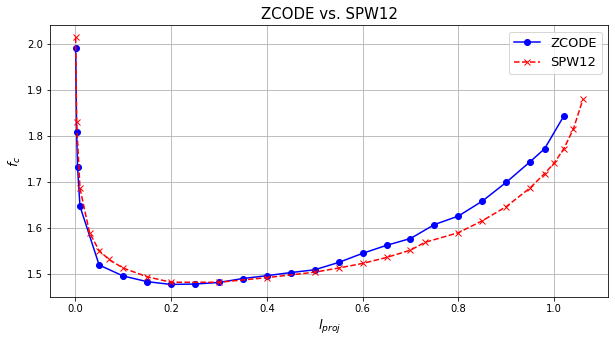}
    \caption{Plot showing outgoing luminosity (in units of the Eddington luminosity) vs. color-correction values, for both ZCODE and SPW12 \cite{Suleimanov_2012}. Comparable line trends and color-correction values indicate the validity of ZCODE's methods.}
    \label{Fig6}
\end{figure}

The color characterizes the spectrum shift and the value of $f_c$ is especially useful for comparing with observations and similar models.
Comparing the resulting color correction factors from our model at different luminosities with those obtained from a model with the same conditions and similar methods (SPW12 \cite{Suleimanov_2012}), we found strong agreement in the general line trend, as can be seen in Fig.\ \ref{Fig6}. We note that comparison between computational models is not validation, but this consistency indicates no serious errors in the method or bugs in the code.

\subsection{A note on performance}

As detailed above, ZCODE cycles iteratively to obtain a static atmosphere solution, and the number of cycles 
required for convergence depends on several factors,
notably the initial guess for the atmosphere. Also, we found
that the computational time per cycle
is a factor of 10 larger for simulations with $N_{\rm tot}
= 1e5$ over $N_{\rm tot} = 1e4$, but that the larger number of Monte Carlo particles yields better convergence. Thus the 
computational cost can vary. ZCODE is parallelized
with MPI and each simulation was run using 16 MPI processes, with one CPU core per process.
We found that a typical $N_{\rm tot} = 1e5$ simulation took approximately 120 hours on 16 cores
of the Intel Xeon Haswell 2.0 GHz CPUs of our campus cluster.

\section{Discussion}

\subsection{Behavior of the Total Outgoing Flux}

\hspace*{2em}Figs.\ \ref{Fig2} and \ref{Fig2a} display the evolution of the total outgoing flux ($F_{\text{tot}}$) as a function of cycle number for $l = 0.5$. We note that these figures may appear to correctly depict the time-dependent flux evolution we would expect in an X-ray burst, characterized by a rapid rise due to burning followed by a gradual cooling phase, the simulation does not explicitly model time evolution but rather models time-independent atmospheric solutions and what we are seeing is convergence of those states. 
Each simulation cycle thus represents a ``snapshot" of hotter and cooler states of the atmosphere,  Because the timescales for radiative (around $1\space\text{ns}$) and hydrostatic (around $1\space \mu \text{s}$) equilibrium are much shorter than the burst duration (around $0.1$ to $10 \space \text{s}$), this quasi-static approach can still provide insight into the kinds of atmospheric states that would occur during a burst.

The left panel of 
Fig.\ \ref{Fig2} showcases the values of $F_{\text{tot}}$ for different values of the minimum optical depth parameter ($\tau_{\text{min}}$).  The differences between curves in the initial stages, as the numerical system adjusts from the initial conditions before reaching a plateau, indicate that $\tau_{\text{min}}$ affects the initial conditions in the simulation. Specifically, it is clear from Fig.\ \ref{Fig2} that larger values of $\tau_{\text{min}}$ result in lower values of the initial guess for the total outgoing flux. 
The parameter $\tau_{\text{min}}$ determines the minimum optical depth considered in the atmosphere model and lower $\tau_{\text{min}}$ values correspond to including more of the outer atmosphere, where deviations from blackbody radiation are most significant. This explains the initial differences in flux values for different $\tau_{\text{min}}$, as the system's initial guess and relaxation process are affected by how much of the outer layers are modeled. This makes physical sense, considering that the outer layers of the atmosphere are more transparent, meaning that radiation can escape more freely, explaining why the initial guess for $F_{\text{tot}}$ is larger for lower values of $\tau_{\text{min}}$. Moreover, the fact that the different $\tau_{\text{min}}$ curves converge to a common final value suggests that, despite differences in the initial conditions, the long-term solution is robust. 

Fig.\ \ref{Fig2a} portraying the outgoing flux as a function of 
cycle number similarly shows the solutions are robust.
The flux adjustments throughout the simulation cycles are smoother as N goes up, particularly in the last few cycles as seen in the detail.

\subsection{The Dilute Blackbody Distribution}

\hspace*{2em}The deviation from a pure blackbody spectrum in  Fig.\ \ref{Fig45combined} arises from frequency-dependent opacity sources in the atmosphere. Unlike an idealized blackbody, where radiation is fully thermalized at all frequencies, real astrophysical environments exhibit varying degrees of absorption and scattering, altering how photons escape.

At low photon energies, free-free absorption (inverse Bremsstrahlung) is dominant, where free electrons interact with ions, absorbing photons. This process leads to higher opacity, or a shorter mean-free-path, causing strong suppression of low-frequency photons in the emergent spectrum.
At intermediate frequencies, photoionization and photoexcitation (bound electrons in atoms absorb photons, causing ionization or excitation) contribute significantly. Although to a lesser extent,  this increases the opacity for these photons as well, meaning radiation struggles to escape, and explaining why the curve in Fig.\ \ref{Fig45combined} appears suppressed and very slightly jagged in this regime.

At high photon energies, scattering effects—Compton scattering and Thomson scattering—become dominant over absorption. In Compton scattering, photons lose energy through interactions with electrons, while Thomson scattering allows photons to escape elastically without losing energy. The net result is that, unlike lower-frequency photons that tend to be absorbed,  higher-frequency photons scatter more easily but are more likely to escape, hence the smoothness of the curve in Fig.\ \ref{Fig45combined} at high energies. Consequently, the effective opacity generally decreases with increasing photon frequency, shifting the spectrum toward higher frequencies, or a ``color temperature" higher than the effective photosphere temperature determined from the Stefan-Boltzmann law, which considers only the total luminosity and stellar radius.

This interplay between absorption and scattering processes fundamentally shapes the outgoing spectrum, leading to the blueshifted appearance observed in Fig.\ \ref{Fig45combined}, showcasing an equilibrium spectrum produced by our model in which the outgoing light curve is blueshifted.

\subsection{Outgoing Spectra at Different Luminosities}

\hspace*{2em}As we can see in the right panel of Fig.\ \ref{Fig45combined}, as the luminosity increases, in the outgoing radiation spectrum the peak flux shifts to higher photon frequency, moving from $4.74$~keV at $l = 0.2$ to $8.46$~keV at $l = 1.02$. In the blackbody approximation, this follows directly from Planck’s law, which states that the peak of the spectrum shifts to higher energies as temperature increases. This trend is consistent with a system where radiation is increasingly dominated by thermal processes as luminosity rises.
Additionally, we can observe that at higher luminosities the overall shape of the spectrum more closely resembles a blackbody distribution, at least qualitatively. 

\subsection{Color-Correction Factors}

\hspace*{2em}We recall that the color correction factor defines the equilibrium, outgoing radiation spectrum's fit to a diluted Planck function, and thus serves a means of comparing our final result to other models or observational data. As aforementioned, the fact that the relationship between the color correction factors and luminosity in Fig.\ \ref{Fig6} aligns well with the results from similar methodologies indicates that our simulation is generally well-behaved.

From \cite{Suleimanov_Poutanen_Werner_2011}, we know that in these X-ray bursting neutron star atmospheres there should generally be a local minimum of the color correction factor at some intermediate luminosity ($l \approx 0.1$--$0.5$). This is clearly the case in Fig.\ \ref{Fig6}, as we notice a minimum for $f_c$ around $l \approx 0.2$. Moreover, \cite{Suleimanov_Poutanen_Werner_2011} also discusses how the decreasing role of Compton scattering compared to ``true opacity" (i.e.\ absorption processes, see \cite{Suleimanov_Poutanen_Werner_2011})  as luminosity increases means that the color-correction factors should increase with luminosity above the minimum. This behavior is also shown in Fig.\ \ref{Fig6},  where we observe an exponential increase in the value of $f_c$ from $l \approx 0.2$--$1.0$ and above. Finally, \cite{Suleimanov_Poutanen_Werner_2011} notes that at the lowest luminosities, although Compton scattering is not very significant, the color-correction factor should increase again due to properties related to the free-free opacity (see \cite{Suleimanov_Poutanen_Werner_2011}), which is again what we see in Fig.\ \ref{Fig6}. Generally, the fact that our results follow the expected features for $f_c$ - $l$ dependencies further verifies the model's methods. Detailed explanations of the scattering and absorption processes governing these behaviors can be found in \cite{Suleimanov_Poutanen_Werner_2011}.

\section{Conclusion}

\hspace*{2em} Overall, we have built confidence in the simulation's methods by thoroughly testing the baseline case. By examining the relationship between the simulation cycle number and the total outgoing flux, we demonstrated how the model numerically calculates through a sequence of time-independent atmospheric ``snapshots", each iteratively refined,  which are used to progressively converge toward the correct atmospheric state during the burst. Furthermore, by analyzing variations in the outgoing flux for different minimum optical depths, we confirmed that the model consistently converges to a common steady-state value, despite minor discrepancies in the initial flux estimate.

Moreover, we explored the meaning and physical explanations for the ``diluted" blackbody distribution we observe as the equilibrium outgoing spectrum. Specifically, we discussed how the observed spectral deviations from a pure blackbody are well explained by the influence of frequency-dependent opacity sources in the atmosphere. The higher opacity for photons at low and intermediate energies in the atmosphere explains both the suppression of the emergent spectrum and the statistical fluctuations in flux in this regime. On the other hand, the dominant effects of scattering at high photon energies explains the enhanced photon escape probability and redistribution of photon energies in this regime.

Additionally, through our analysis of spectra at different luminosities we confirmed the expected trend that increasing luminosity results in a shift of the peak emission toward higher photon energies. The closer qualitative resemblance to a blackbody spectrum at higher luminosities also suggests that as luminosity increases, the system becomes increasingly thermalized.

Finally, and most significantly for the purpose of testing the model, we showed very strong consistency in our model's $f_c$ - $l$ relationship with a previous model and theoretical expectations. This builds confidence in our methodology and suggests that different model configurations can and should be tested to explore new parameter spaces.  Importantly, the convergence of the results was a critical factor in achieving this consistency. 
Poor statistical sampling at certain frequencies or atmospheric regions led to errors, which we mitigated by increasing the particle count. Future work would benefit from a more efficient sampling scheme that targets regions and frequencies that require more precise sampling.

\ack
This research was supported in part by the U.S. Department of Energy (DOE) 
under grant DE-FG02-87ER40317. 
The authors would like to thank Stony Brook Research Computing and Cyberinfrastructure, and the Institute for Advanced Computational Science at Stony Brook University for access to the high-performance SeaWulf computing system, which was made possible by a \$1.4M National Science Foundation grant (\#1531492).
This work was performed in part under the auspices of the Los Alamos
National Laboratory, operated by Triad National Security, LLC for
the National Nuclear Security Administration of the DOE under Contract No. 89233218CNA000001. 

%\section*{References}

\bibliography{references}

\end{document}